\documentclass[12pt,twoside]{article}
\usepackage{epsfig,float,wrapfig}
\usepackage{color}
                                \usepackage{fancyhdr}

\usepackage{amssymb,amsmath}

\newcommand{\be}{\begin{equation}}
\newcommand{\ba}{\begin{eqnarray}}
\newcommand{\ea}{\end{eqnarray}}

\def\a{\alpha}
\def\b{\beta}

\def\d{\delta}
\def\e{\epsilon}

\def\f{\phi}

\def\g{\gamma}

\def\l{\lambda}
\def\m{\mu}
\def\n{\nu}

\def\p{\pi}
\def\q{\theta}
\def\r{\rho}
\def\s{\sigma}
\def\t{\tau}

\def\x{\xi}

\def\D{\Delta}

\def\OO{\Omega}
\def\P{\Pi}

\def\S{\Sigma}

\def\ca{{\cal A}}

\def\cs{{\cal S}}

\def\cw{{\cal W}}

\newcommand{\ti}{\tilde}

\newcommand{\pa}{\partial}

\newcommand{\chat}{\bar{c}}

\newfloat{vertex}{h}{top}
\newfloat{prop}{h}{top}

\renewcommand{\title}[1]{\null\vspace{25mm}\noindent{\Large{\bf #1}}\vspace{10mm}}
\newcommand{\authors}[1]{\noindent{\large #1}\vspace{20mm}}
\newcommand{\address}[1]{{\center{\noindent #1\vspace{10mm}}}}
\renewcommand{\abstract}[1]{\vspace{17mm}
\noindent{\small{\em Abstract.} #1}\vspace{2mm}}     
 
\begin{document}   \setcounter{table}{0}
 
\begin{titlepage}
\begin{center}
\hspace*{\fill}{{\normalsize \begin{tabular}{l}
                              {\sf hep-th/0004071}\\
			      {\sf REF. TUW 00-11}\\
			      {\sf \today}\\
                              \end{tabular}   }}

\title{Perturbative Chern-Simons Theory on \\ Noncommutative $\mathbb{R}^{3}$}

\authors {  \Large{ A. A. Bichl, J. M. Grimstrup$^{1}$, V. Putz, M. Schweda$^{2}$} }    \vspace{-20mm}
       
\address{Institut f\"ur Theoretische Physik, Technische Universit\"at Wien\\
      Wiedner Hauptstra\ss e 8-10, A-1040 Wien, Austria}
\footnotetext[1]{Work supported by ``The Danish Research Academy''.}
\footnotetext[2]{Work supported by the ``Fond zur F\"orderung der Wissenschaftlichen Forschung'' under Project Grand Number P11582-PHY.}       
\end{center} 
\thispagestyle{empty}

\abstract{
A $U(N)$ Chern-Simons theory on noncommutative $\mathbb{R}^{3}$ is constructed as a $\q$-deformed field theory. The model is characterized by two symmetries: the BRST-symmetry and the topological linear vector supersymmetry. It is shown that the theory is finite and $\q_{\m\n}$-independent at the one loop level and that the calculations respect the restriction of the topological supersymmetry. Thus the topological $\q$-deformed Chern-Simons theory is an example of a model which is non-singular in the limit $\q \rightarrow 0$.                  }


\end{titlepage}


\pagenumbering{arabic}

\newcommand{\xh}{\widehat{x}}

\clearpage
\section{Introduction}
It is well known that at the Planck scale the flat space-time must be modified drastically leading to a noncommutative space-time. The corresponding field theories have to be formulated in the framework of noncommutative geometry \cite{Doplicher:1994zv}.\\
In this short letter we discuss the Chern-Simons theory with respect to a non-Abelian $U(N)$-gauge group on noncommutative $\mathbb{R}^{3}$. In the corresponding $\q$-deformed field theory we are able to establish the usual BRST-symmetry \cite{Becchi:1974xu} and the topological vector supersymmetry (VSUSY) \cite{Piguet:1995er}. The vector-like generators of this VSUSY give rise together with the BRST-operator to a Wess-Zumino-like anticommutation relation which closes on-shell on the space-time translations.\\
At the classical level the Ward-identity characterizing the VSUSY is linearly broken in the quantum fields. These breakings are induced by the external sources needed for the description of the nonlinear terms of the BRST-transformations.\\
It is well known that in the commutative case the perturbative finiteness of the Chern-Simons model is governed by the VSUSY within the algebraic renormalization procedure \cite{Piguet:1995er}.\\
In the presented notes we are able to show that the Chern-Simons model is finite and independent of the deformation parameter $\q_{\m\n}$ at least at the one loop level. Additionally, the perturbative calculations are in agreement with the restrictions coming from the VSUSY.\\
The letter is organized as follows. Section 2 gives a short presentation of $\q$-deformed field theory. In section 3  we present the Chern-Simons theory as a $\q$-deformed field theory on a noncommutative $\mathbb{R}^{3}$. The BRST-transformation and the VSUSY are introduced at the tree level. Section 4 is devoted to sketch the perturbative calculation at the one loop level.
\section{$\q$-deformed Field Theory}
The noncommutative $\mathbb{R}^{3}$ is defined as the algebra $\ca_{x}$ generated by $\xh_{\mu}$, $\mu=1,2,3$ satisfying the commutation relation \cite{Filk:1996dm}
\be
[\xh_{\m},\xh_{\n}] = {\rm{i}}  \q_{\m\n},
\end{equation}
where $\q_{\m\n}$ is a real, constant, antisymmetric matrix with rank 3. For a function $f(x)$ on ordinary $\mathbb{R}^{3}$ one associates an element $W(f)$ of the algebra $\ca_{x}$ by 
\be
W(f)=\frac{1}{(2\p)^{\frac{3}{2}}}\int d^{3}k \,e^{{\rm{i}}k_{\m}\xh_{\m}}\ti{f}(k),\label{eq.1}
\end{equation}
where $\ti{f}(k)$ is the Fourier transform of $f(x)$
\be
\ti{f}(k)=\frac{1}{(2\p)^{\frac{3}{2}}}\int d^{3}x\, e^{-{\rm{i}}k_{\m}x_{\m}}f(x).
\end{equation}
Using relation (\ref{eq.1}) the star product on $\mathbb{R}^{3}$ is defined by 
\be
W(f)W(g)=W(f\star g),\label{eq.2}
\end{equation}
and one finds 
\be
(f\star g)(x)=\int \frac{d^{3}k }{(2\p)^{ \frac{3}{2} }}\int \frac{ d^{3}p }{(2\p)^{ \frac{3}{2} }} \,e^{{\rm{i}}(k_{\m}+p_{\m})x_{\m}}e^{-\frac{{\rm{i}}}{2}\q_{\m\n}k_{\m}p_{\n}}\ti{f}(k)\ti{g}(p).\label{ny-1}
\end{equation}
Relations (\ref{eq.1}) and (\ref{eq.2}) render a representation of functions on the algebra $\ca_{x}$ \cite{Madore:2000en}, and thus allow us to examine gauge theory on $\ca_{x}$ by considering the counterpart on ordinary $\mathbb{R}^{3}$ with the usual product replaced by the star product. 
From (\ref{ny-1}) follows
\be
\int d^{3}x (f\star g)(x)= \int d^{3}x f(x)g(x). \label{BILIN}
\end{equation}
Similarly, one gets for a triple product
\ba
\left(f\star g\star h\right) (x) &=& \int \frac{d^{ 3}p_{1}}{(2\p)^{ \frac{3}{2} }} 
                                          \frac{d^{ 3}p_{2}}{(2\p)^{ \frac{3}{2} }}
                                          \frac{d^{ 3}p_{3}}{(2\p)^{ \frac{3}{2} }}     
 \;e^{-\frac{{\rm{i}}}{2}\q_{\m\n}\left[(p_{1})_{\m}(p_{2})_{\n} + (p_{1})_{\m}(p_{3})_{\n} +  (p_{2})_{\m}(p_{3})_{\n}\right]} \nonumber\\
 && \times \, e^{{\rm{i}} (p_{1} + p_{2} + p_{3})_{\m}x_{\m}} \ti{f}(p_{1})\ti{g}(p_{2})\ti{h}(p_{3}).
\ea
With $n$ arbitrary fields $\f_{k}$ one can show the validity of cyclic permutations
\ba
\int_{\mathbb{R}^{3}} d^{3}x \left(\f_{1}\star\f_{2}\star \cdots \star\f_{n}\right)(x) &=& (-1)^{g_{1}(g_{2} + \,\cdots\, +g_{n} )}\int_{\mathbb{R}^{3}} d^{3}x\left(\f_{2}\star\cdots \star\f_{n}\star\f_{1}\right) (x),\label{eq.2a}
\ea
where $g_{i}$ is the total grading of the field $\f_{i}$. Additionally, one can show that the functional differentiation holds too, {\it i.e.}
\ba
\frac{\d}{\d \f_{1}(y)} \int_{\mathbb{R}^{3}} d^{3}x  \left(\f_{1}\star\f_{2}\star \cdots \star\f_{n}\right)(x) &=& \left(\f_{2}\star \cdots \star\f_{n}\right)(y).\label{eq.2b}
\ea
Equations (\ref{eq.2a}) and (\ref{eq.2b}) imply now that the off-shell algebra of the relevant functional operators describing the symmetry content of the model remains valid.
\section{BRST-symmetry and Linear VSUSY}
With a Lie-algebra valued gauge field $A_{\m}(x)=A^{a}_{\m}(x)T^{a}$, where $T^{a}$ are the $N^{2}$ generators of the U($N$)-group,\footnote{Not all gauge groups are realizable in noncommutative field theory \cite{Madore:2000en,Matsubara:2000gr,Terashima:2000xq}} $a=0,1,...,(N^{2}-1)$
\be
\left(T^{0}\right)_{mn}=\left(\frac{2}{N}\right)^{\frac{1}{2}}\d_{mn}, \,\,\,\,\,\left(T^{a} \right)_{mn}\nonumber
\end{equation}
with
\be
[T^{a},T^{b}]= 2{\rm{i}}f^{abc}T^{c}, \,\,\,\,\,
\{T^{a},T^{b}\} =  2d^{abc}T^{c},\,\,\,\,\, \mbox{Tr}\left( T^{a}T^{b} \right)= 2 \d^{ab} ,  \label{GROUP}
\end{equation}
one can define the gauge-invariant classical metric independent action as
\be
\S_{cl} = - \int d^{3}x\;\frac{1}{2}\e_{\m\n\r}\;\mbox{Tr}\left(A_{\m} \pa_{\n}A_{\r} - \frac{2{\rm{i}}}{3}A_{\m}\star A_{\n} \star A_{\r} \right), \label{ACTION-cl}
\end{equation}
which is invariant under the infinitesimal gauge transformation
\be
\d A_{\m} = \pa_{\m}\l + {\rm{i}} \left( \l\star A_{\m} -A_{\m}\star\l   \right) \equiv D_{\m}\l,
\end{equation}
where $\l$ is a Lie-algebra valued gauge parameter. The corresponding field strength is given by
\be
F_{\m\n} = \pa_{\m}A_{\n}-\pa_{\n}A_{\m} - {\rm{i}} \left( A_{\m} \star A_{\n} - A_{\n}\star A_{\m} \right).
\end{equation}
In order to quantize the model within the BRST-scheme, the gauge-symmetry is replaced by the nilpotent BRST-symmetry \cite{Becchi:1974xu,Piguet:1995er}
\be
sA_{\m} = D_{\m}c,\,\,\,\,\,\,
sc = {\rm{i}} c\star c,
\end{equation}
where $c$ is the anticommuting Faddeev-Popov ghost field. Within the quantization procedure a BRST-invariant gauge-fixing must be introduced in the following manner
\be
\S_{gf} = s\; \int d^{3}x \;\mbox{Tr}\left( \bar{c}\star \pa_{\m} A_{\m} \right),
\end{equation}
with
\be
s\chat = B ,\,\,\,\,\, sB =0,
\end{equation}
where $\bar{c}$ is the antighost field and $B$ the multiplier field implementing the Landau gauge. 
The gauge fixing part of the action of course depends on the metric, chosen here as the flat Euclidian one $\d_{\m\n}$. 
The total action is now
\be
\S = \S_{cl} + \S_{gf}.\label{ACTION-gf}
\end{equation}
Besides the BRST-invariance, (\ref{ACTION-gf}) possesses an additional global supersymmetry, whose generators carry the Lorentz index \cite{Piguet:1995er}.
\ba
\d_{\n} A_{\t}  &=&  \e_{\m\n\t}\pa_{\m}\bar{c},\nonumber\\
\d_{\n} c       &=&  - A_{\n},\nonumber\\
\d_{\n} B       &=&  - \pa_{\n}\chat,\nonumber\\
\d_{\n} \chat   &=& 0.
\ea
The operator $\d_{\n}$ gives rise together with the BRST-operator to the anticommutation relations
\ba
  \d_{\n} \d_{\t} + \d_{\t} \d_{\n}  &=& 0, \\
\d_{\n}  s + s  \d_{\n} &=& - \pa_{\n} + \mbox{(eq. of motion)},
\ea
which close on-shell on space time translations.
In order to describe the symmetry content of the model with respect to the BRST-symmetry and the linear VSUSY, one has to write down the corresponding Ward-identities (WI) at the classical level. In order to carry out this procedure, one introduces external unquantized sources for the nonlinear pieces of the BRST-transformations
\be
\S_{ext} = \int d^{3}x \;\mbox{Tr}\left( \r_{\m} \star sA_{\m} + \s \star sc \right).
\end{equation}
Then the total corresponding tree-level action is given by
\ba
\S &=& \S_{cl} + \S_{gf} + \S_{ext}\nonumber\\
 &=&- \int d^{3}x\;\frac{1}{2}\e_{\m\n\r}\;\mbox{Tr}\left(A_{\m} \pa_{\n}A_{\r} - \frac{2{\rm{i}}}{3}A_{\m}\star A_{\n} \star A_{\r} \right) \nonumber\\
 &+& \int d^{3}x \;\mbox{Tr}\left( B  \pa_{\m}A_{\m} - \bar{c}\star \pa_{\m} D_{\m}c \right)\nonumber\\
&+&  \int d^{3}x \;\mbox{Tr}\left( \r_{\m} \star D_{\m}c + {\rm{i}}\s\star  c\star c \right).\label{ACTION-full}
\ea
The total action satisfies the nonlinear, $\q$-deformed Slavnov-Taylor identity
\be
\cs\left( \S \right) =0,
\end{equation}
where
\be
\cs\left( \S \right) = \int d^{3}x \;\mbox{Tr}\left( \frac{\d \S}{\d \r_{\m}}\star \frac{\d\S}{\d A_{\m}} +  \frac{\d \S}{\d \s} \star \frac{\d\S}{\d c} +    B \star \frac{\d\S}{\d \chat}      \right),\label{ST}
\end{equation}
which describes the BRST-symmetry. The VSUSY is characterized by the following integrated functional differential operator
\ba
\cw_{\m}&=& \int d^{3}x \;\mbox{Tr}\left(- \e_{\n\m\t}\left( \pa_{\n}\chat + \r_{\n} \right) \star \frac{\d}{\d A_{\t}} - A_{\m} \star \frac{\d}{\d c} - \pa_{\m} \chat \star \frac{\d}{\d B} - \s\star\frac{\d}{\d \r_{\m}}\right).\label{SUSY}
\ea
The WI operator (\ref{SUSY}) yields the linearly broken supersymmetry
\be
\cw_{\m} \S = \D_{\m}^{cl},\label{VSUSY-WI}
\end{equation}
with
\be
\D_{\m}^{cl} = \int d^{3}x \;\mbox{Tr}\left( -\e_{\n\m\t}\r_{\n}\pa_{\t}B + \r_{\t}\pa_{\m}A_{\t} - \s\pa_{\m}c\right),
\end{equation}
where $\D_{\m}^{cl}$ is linear in the quantum fields.
\section{One-loop Calculations}
In order to check the one-loop UV and IR behaviour of the noncommutative Chern-Simons model, one needs the corresponding Feynman rules. For the various propagators of the model only the bilinear part of the full action (\ref{ACTION-full}) is relevant. However, due to the fact that noncommutativity does not affect bilinear parts of the action (\ref{BILIN}), the propagators are the usual ones \cite{Guadagnini:1989kr}. In momentum space one therefore has

\begin{wrapfigure}[2]{l}{2cm}
\epsfig{figure=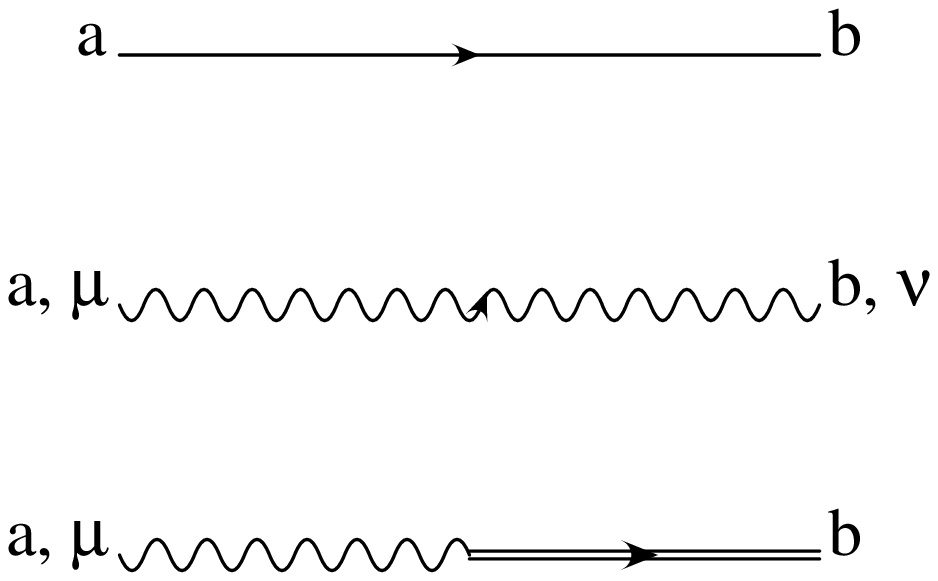,height=3.3cm,width=4.5cm}
\end{wrapfigure}
\ba
\ti{G}^{c\chat,ab}(k)       &=&    \frac{\d^{ab}}{(2\p)^{\frac{3}{2}}}        \frac{1}{k^{2}},\\
\ti{G}^{AA,ab}_{\m\n}(k) &=&    \frac{ {\rm{i}} \d^{ab}}{(2\p)^{\frac{3}{2}}}  \frac{\e_{\m\n\l}k_{\l}}{k^{2}},\\
\ti{G}^{AB,ab}_{\m}(k)      &=&-    \frac{{\rm{i}} \d^{ab}}{(2\p)^{\frac{3}{2}}}   \frac{k_{\m}}{k^{2}}.
\ea
The interaction piece of the action (\ref{ACTION-full}) determines the corresponding Feynman-rules for the three gluon vertex and the ghost-antighost-gluon vertex:
\ba
\S_{int} &=& \int d^{3}x \;\mbox{Tr}\left( \frac{{\rm{i}} }{3}\e_{\m\n\r} A_{\m}\star A_{\n}\star A_{\r} + {\rm{i}} \pa_{\m}\chat \star \left(  c\star A_{\m} - A_{\m}\star c  \right)\right) \nonumber\\
 &=&\mbox{Tr}^{abc} \int d^{3}x \left( \frac{{\rm{i}} }{3}\e_{\m\n\r} A^{a}_{\m}\star A^{b}_{\n}\star A^{c}_{\r}  + {\rm{i}} \pa_{\m}\chat^{a} \star c^{b}\star A^{c}_{\m} - {\rm{i}} \pa_{\m}\chat^{a} \star A^{b}_{\m}\star c^{c}\right),
\ea
where $\mbox{Tr}^{abc} =\mbox{Tr}\left(T^{a}T^{b}T^{c}\right)$. From (\ref{GROUP}) follows
\be
\mbox{Tr}^{abc} = 2 \left( {\rm{i}} f^{abc} + d^{abc} \right),
\end{equation}
and therefore the corresponding Feynman rules for the interaction vertices in momentum space become 
\begin{wrapfigure}[8]{l}{2cm}
\epsfig{figure=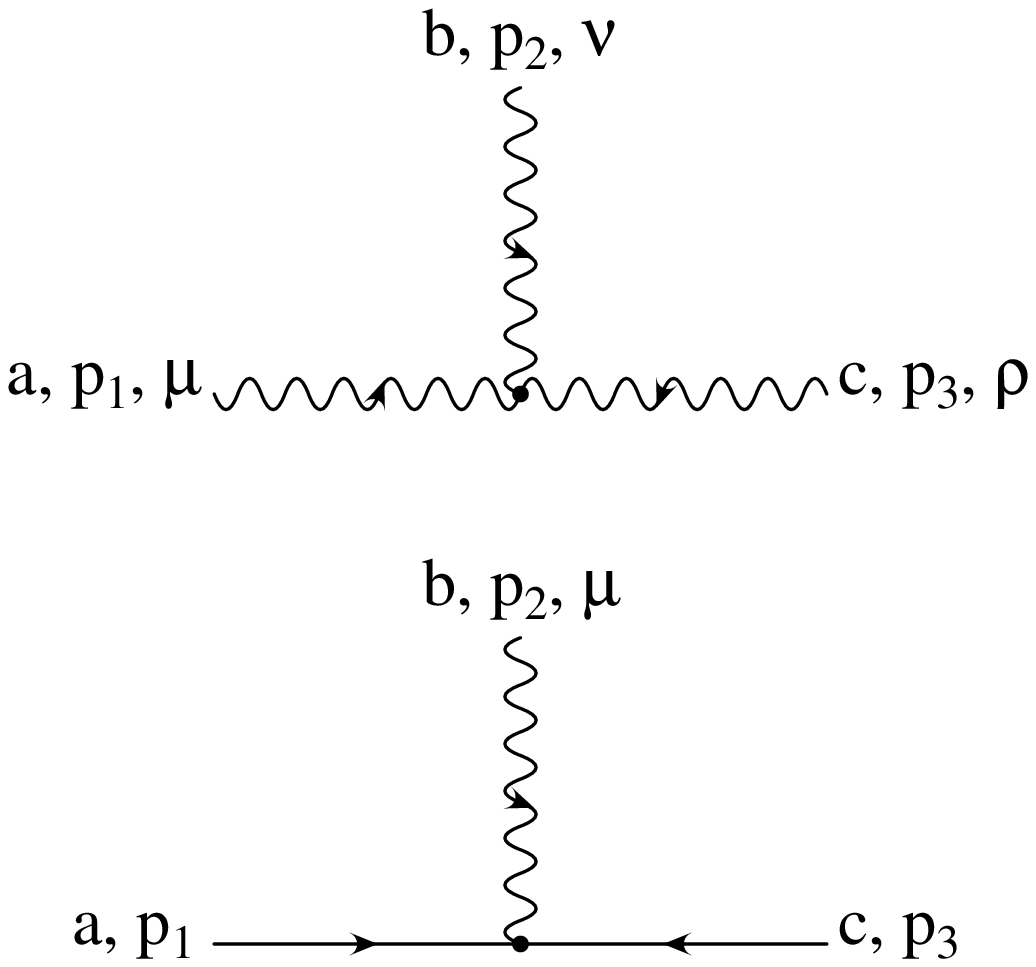,width=5cm}
\end{wrapfigure}
%
%
%
%
%
%
 $  $   \\

\ba
&&\ti{V}^{AAA,abc}_{\m\n\r}(p_{1},p_{2},p_{3})=  \frac{1}{(2\p)^{\frac{3}{2}}}  \; \e_{\m\n\r}\OO^{abc}_{\q}(p_{2},p_{3}),\label{4}\\
\mbox{}\nonumber\\
\mbox{}\nonumber\\
\mbox{}\nonumber\\
&&\ti{V}^{cA\chat,abc}_{\m}(p_{1},p_{2},p_{3})= -\frac{1}{(2\p)^{\frac{3}{2}}} {\rm{i}} \;(p_{3})_{\m}\OO^{abc}_{\q}(p_{2},p_{3}),\label{5}
\ea

$  $   \\

\noindent where $\OO^{abc}_{\q}(p_{2},p_{3})$ is defined by
\be
\OO^{abc}_{\q}(p_{2},p_{3}) = 4 \left[f^{abc}\mbox{cos}\left(\frac{1}{2}p_{2}\times p_{3}\right) - d^{abc}\mbox{sin}\left(\frac{1}{2}p_{2}\times p_{3}\right)\right],\label{OMEGA}
\end{equation}
with $p_{2}\times p_{3}=\q_{\m\n}(p_{2})_{\m}(p_{3})_{\n}$.
In order to carry out explicit calculations the following contractions of the structure constants $f^{abc}$ and $d^{abc}$ are useful
\ba
f^{abc}f^{dbc} &=& N(\d^{ad} - \d^{a0}\d^{d0})\label{rel1},\\
d^{abc}d^{dbc} &=& N(\d^{ad} + \d^{a0}\d^{d0}).\label{rel2}
\ea
From the commutation relations we have additionally,
\ba
f^{abc} &=& \frac{1}{4 {\rm{i}} } \mbox{Tr}\left( \left[T^{a},T^{b}\right]T^{c}\right),\nonumber\\
d^{abc} &=& \frac{1}{4} \mbox{Tr}\left( \left\{T^{a},T^{b}\right\}T^{c}\right).
\ea
In order to verify the relations (\ref{rel1}) and (\ref{rel2}) one has to use the following completeness relation for generators $T^{a}$ of the $U(N)$ gauge group
\be
\left(T^{a}\right)_{mn}\left(T^{a}\right)_{m'n'}= 2 \d_{mn'}\d_{nm'}, \,\,\,\,\,\,\,\,\,\,\,\, m,n=1,...,N   .
\end{equation}
Before doing explicitly the one loop analysis we want to stress that already at the tree level the consequences of the VSUSY WI are fulfilled. In doing a Legendre transformation one gets from the WI (\ref{VSUSY-WI}) that the propagator of the gluon field and the ghost-antighost-field propagator are related. In momentum space one obtains
\be
\e_{\m\n\l}{\rm{i}}k_{\l}\ti{G}^{c\chat,ab}(k) = \ti{G}^{AA,ab}_{\m\n}(k).
\end{equation}
For the three point vertices the following WI follow directly from (\ref{VSUSY-WI})
\be
{\rm{i}}\left(p_{3}\right)_{\n}\e_{\m\n\t}\ti{V}^{AAA,abc}_{\l\s\t}(p_{1},p_{2},p_{3})=\d_{\m\s}\ti{V}^{cA\chat,abc}_{\l}(p_{1},p_{2},p_{3}) - \d_{\m\l}\ti{V}^{cA\chat,abc}_{\s}(p_{1},p_{2},p_{3}).
\end{equation}
These relations are indeed responsible for the one loop perturbative finiteness. In fact, one finds that the gluon loop and the ghost loop contribution to the gluon self energy $\P^{ab}_{\m\n}$ are equal up to a sign, and thus cancel. A similar result is found for the one loop vertex corrections which therefore also vanish. Therefore, the loop cancellation present in the ordinary Chern-Simons model \cite{Guadagnini:1989kr} persists in the noncommutative case, {\it i.e.} the noncommutative Chern-Simons model is finite to first loop order.\\
In the following we will explicitely perform the one loop analysis of the gluon self energy graph (a) of fig.1 and show that the 'non-planar' contribution arising due to the noncommutativity is finite. Using the above Feynman rules one calculates the following expression for the gluon self-energy with an internal gluon field,
\begin{figure}[t]
\begin{center}
\epsfig{figure=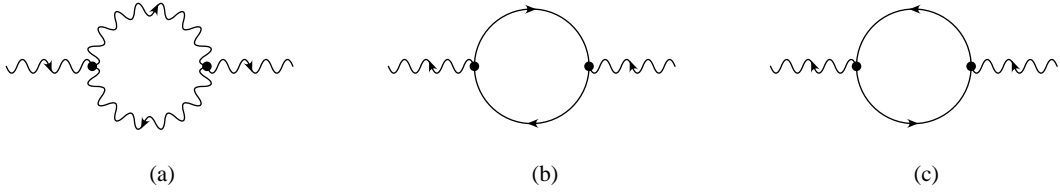}
\caption{gluon self energy}
\end{center}
\end{figure}
\be
\P^{ab}_{\m\n}(p,-p)= - \frac{1}{(2\p)^{6}}\int \frac{d^{3}k}{(2\p)^{\frac{3}{2}}} \frac{\left(k_{+}\right)_{\m}\left(k_{-}\right)_{\n} +\left(k_{+}\right)_{\n}\left(k_{-}\right)_{\m}      }{k^{2}_{+}k^{2}_{-}}\OO^{acd}_{\q}(-k_{+},k_{-})\OO^{cbd}_{\q}(-p,-k_{-}),
\end{equation}
where $k_{+}= k + \frac{p}{2}$ and $k_{-}= k - \frac{p}{2}$. Using now (\ref{rel1}) and (\ref{rel2}) one gets
\be
\P^{ab}_{\m\n}(p,-p)=  \frac{16N}{(2\p)^{6}}\int  \frac{d^{3}k}{(2\p)^{\frac{3}{2}}}     \frac{\left(k_{+}\right)_{\m}\left(k_{-}\right)_{\n} +\left(k_{+}\right)_{\n}\left(k_{-}\right)_{\m}      }{k^{2}_{+}k^{2}_{-}}  \left[ \d^{ab} - \d^{a0}\d^{b0}\frac{1}{2} \left( e^{{\rm{i}} p\times k} + e^{-{\rm{i}} p \times k }  \right) \right]. \label{hurra}
\end{equation}
The first 'planar' term is just the result which corresponds to the commutative model. From the result (\ref{hurra}) it is also seen that one has to use a $U(N)$ gauge group in order to have a nontrivial result. Now one can use the result of \cite{Guadagnini:1989kr} with an appropriate regularization scheme in order to discuss the $\q$-independent expression of eq. (\ref{hurra}). The further discussion of the $\q$-dependent, 'non-planar' part of (\ref{hurra}) can be done with the techniques presented already in \cite{Hayakawa:1999zf}.\\
Using Schwinger parametrization
\be
\frac{1}{k_{+}^{2}}= \int^{\infty}_{0}d\a_{+}e^{-\a_{+}k_{+}^{2}}
\end{equation}
we discuss now
\be
\P'_{\m\n}(p,-p)= - \int \frac{d^{3}k}{(2\p)^{\frac{3}{2}}} \frac{\left(k_{+}\right)_{\m}\left(k_{-}\right)_{\n} +\left(k_{+}\right)_{\n}\left(k_{-}\right)_{\m}      }{k^{2}_{+}k^{2}_{-}} \left( e^{{\rm{i}} p\times k} + e^{-{\rm{i}} p \times k }  \right)  ,
\end{equation}
%
%
%
and get
\ba
\P'_{\m\n}(p,-p)&=&  - \int  \frac{d^{3}k}{(2\p)^{\frac{3}{2}}} \int_{0}^{\infty}d\a_{+}\int_{0}^{\infty}d\a_{-}    \left(  \left(k_{+}\right)_{\m}\left(k_{-}\right)_{\n} +\left(k_{+}\right)_{\n}\left(k_{-}\right)_{\m}    \right) e^{-\a_{+}k^{2}_{+} -\a_{-}k^{2}_{-} } \nonumber\\&& \times\,   \left( e^{{\rm{i}}p\times k} + e^{-{\rm{i}} p \times k} \right).\label{hurra1}
\ea
In defining $\ti{p}_{\m}=p_{\l}\q_{\l\m}$ and
\be
I_{\m\n}(\eta,p)= \left.\left\{\left( \frac{\pa}{\pa y_{\m}} \frac{\pa}{\pa z_{\n}} + \frac{\pa}{\pa y_{\n}} \frac{\pa}{\pa z_{\m}} \right)
\int \frac{d^{3}k}{(2\p)^{\frac{3}{2}}}e^{-\a_{+}\left( k + \frac{p}{2}\right)^{2} -\a_{-}\left( k - \frac{p}{2}\right)^{2} + {\rm{i}}\eta\ti{p}k - y  \left( k + \frac{p}{2}\right) - z  \left( k - \frac{p}{2}\right)      } \right\} \right|_{y=z=0}.\label{nr.49}
\end{equation}
Eq. (\ref{hurra1}) can be written as
\be
\P'_{\m\n}(p,-p)=  -\int_{0}^{\infty}d\a_{+}\int_{0}^{\infty}d\a_{-} \left(   I_{\m\n}(1,p) +  I_{\m\n}(-1,p)\right) .\label{witten}
\end{equation}
In order to carry out the $k$-integration one introduces new variables
\be
k'_{\l} = k_{\l} + \frac{B_{\l}}{\a_{+} + \a_{-}} \equiv k_{\l} + \frac{B_{\l}}{\b},
\end{equation}
with
\be
B_{\l}= \frac{1}{2}\left[ \left(\a_{+} - \a_{-}\right)p_{\l} - {\rm{i}}\eta\ti{p}_{\l} + \left( y + z\right)_{\l}   \right],
\end{equation}
in order to use gaussian integration
\be
\int\frac{d^{3}k'}{(2\p)^{\frac{3}{2}}}e^{-\b {k'}^{2}}= \frac{1}{(2\b)^{\frac{3}{2}}}.
\end{equation}
In this way one gets for (\ref{nr.49})
\be
I_{\m\n}(\eta,p)= \left.\left\{ \left(\frac{\pa}{\pa y_{\m}} \frac{\pa}{\pa z_{\n}} + \frac{\pa}{\pa y_{\n}} \frac{\pa}{\pa z_{\m}} \right)
\frac{1}{(2\b)^{\frac{3}{2}}}e^{\frac{B^{2}}{\b} -\frac{\b p^{2}}{4} - (y-z)\frac{p}{2}   }\right\} \right|_{y=z=0}.
\end{equation}
Doing explicitly the differentiation one finds
\be
I_{\m\n}(\eta,p)= \frac{1}{(2\b)^{\frac{3}{2}}}  
e^{-\frac{\a_{+}\a_{-}}{\b}p^{2} - \frac{\eta^{2}}{4\b}\ti{p}^{2} }\left\{  \frac{1}{\b}\d_{\m\n} -2 \left( \frac{\a_{+}\a_{-}}{\b^{2}}p_{\m}p_{\n} + \eta^{2}\frac{\ti{p}_{\m}\ti{p}_{\n}}{4\b^{2}}  \right) + \mbox{terms linear in $\eta$} \right\},
\end{equation}
implying the following result
\be
\P'_{\m\n}(p,-p)=  -2 \int_{0}^{\infty}d\a_{+}\int_{0}^{\infty}d\a_{-} e^{-\frac{\a_{+}\a_{-}}{\b}p^{2} -\frac{\ti{p}^{2}}{4\b}  } \frac{1}{(2\b)^{\frac{3}{2}}} 
\left\{  \frac{\d_{\m\n}}{\b} - 2 \left( \frac{\a_{+}\a_{-}}{\b^{2}}p_{\m}p_{\n}   + \frac{\ti{p}_{\m}\ti{p}_{\n}}{2\b^{2}} \right) \right\} \label{witten2}   .
\end{equation}
It remains to study the parametric integration (\ref{witten2}). One has two different types of integrals (with $\a=1,2$)
\be
I^{(i)}_{\a} = \int_{0}^{\infty}d\a_{+}\int_{0}^{\infty}d\a_{-}\frac{1}{\b^{\frac{3}{2}+\a}}e^{-\frac{\a_{+}\a_{-}}{\b}p^{2} - \frac{\ti{p}^{2}}{4\b}}.
\end{equation}
\be
I^{(ii)}   =      \int_{0}^{\infty}d\a_{+}    \int_{0}^{\infty}d\a_{-}    \frac{\a_{+} \a_{-} }{\b^{\frac{7}{2}} }   e^{-\frac{\a_{+}\a_{-}}{\b}p^{2} - \frac{\ti{p}^{2}}{4\b}}.\label{second!}
\end{equation}
With the reparametrization
\ba
\a_{+}&=& \x\l    ,\nonumber\\
\a_{-}&=& (1-\x)\l ,\nonumber\\
\r    &=& \x (1-\x )p^{2}\l,
\ea
one arrives at
\be
I^{(i)}_{\a}= \int^{1}_{0}d\x \left( \x(1-\x)p^{2} \right)^{\left(\a -\frac{1}{2}\right)}\int^{\infty}_{0}d\r\frac{1}{\r^{\left(\frac{1}{2}+\a\right)}} e^{-\left(\r + \frac{\x(1-\x)p^{2}\ti{p}^{2}}{4\r}   \right)},
\end{equation}
and
\be
I^{(ii)} = \left( \frac{1}{p^{2}} \right)^{\frac{1}{2}} \int^{1}_{0}d\x \left[ \x(1-\x) \right]^{\frac{1}{2}} \int^{\infty}_{0}\frac{d\r}{\r^{\frac{1}{2}}}e^{-\left( \r + \frac{\x(1-\x)p^{2}\ti{p}^{2}}{4\r}  \right)}
\end{equation}
The $\r$-integration is carried out with the formula \cite{grad}
\be
\int_{0}^{\infty} \frac{dx}{x^{1-\n}}e^{-\g x - \frac{\b}{x}}= 2\left(\frac{\b}{\g}\right)^{\frac{\n}{2}}K_{\n}\left( 2\sqrt{\b\g} \right),
\end{equation}
where $K_{\n}$ is the modified Bessel function \cite{grad,lebe}. In this way one obtains
\be
I^{(i)}_{\a}=2 \left( \frac{4p^{2}}{\ti{p}^{2}} \right)^{ \left( \a - \frac{1}{2} \right)\frac{1}{2}}  \int^{1}_{0}d\x \left[ \x(1-\x) \right]^{\left( \a- \frac{1}{2} \right)\frac{1}{2}}  K_{\frac{1}{2} -\a}\left( z \right) \label{nr.60}.
\end{equation}
where $z=(\x(1-\x)p^{2}\ti{p}^{2})^{\frac{1}{2}} $. Additionally, one has
\be
I^{(ii)} = \left( \frac{1}{p^{2}}\right)^{\frac{1}{2}}\sqrt{2} \int^{1}_{0} d\x \left[ \x(1-\x) \right]^{\frac{1}{2}}\left[ \x(1-\x)p^{2}\ti{p}^{2}  \right]^{\frac{1}{4}}K_{\frac{1}{2}}(z).
\end{equation}
With
\be
K_{\frac{1}{2}}(z)=K_{-\frac{1}{2}}(z)= \left(\frac{\p}{2 z}  \right)^{\frac{1}{2}}    e^{-z},
\end{equation}
and
\be
K_{\n -1}(z) - K_{\n +1}(z) = \frac{-2\n}{z}K_{\n}(z),
\end{equation}
one can show that $I^{(i)}_{1}$ behaves as
\be
I^{(i)}_{1} \sim \left( \frac{1}{\ti{p}^{2}} \right)^{\frac{1}{2}} \times \mbox{finite},
\end{equation}
which shows the usual IR-singularity for $\ti{p}\rightarrow 0$. Whereas $I^{(i)}_{2}$ develops two contributions of the form
\be
I^{(i)}_{2} \sim \left(p^{2}\right)^{\frac{1}{2}} \left( \frac{1}{\ti{p}^{2}} \right) \times \mbox{finite} + \left( \frac{1}{\ti{p}^{2}} \right)^{\frac{3}{2}} \times \mbox{finite}.\label{sixtysix}
\end{equation}
These last terms must be multiplied by $\ti{p}_{\m}\ti{p}_{\n}$. Thus the second term of (\ref{sixtysix}) reproduces an IR-singularity for $\ti{p}\rightarrow 0$, whereas the first term leads to IR-regular, $\q$-independent matrix elements.\\
A similar calculation for $I^{(ii)}$ yields
\be
I^{(ii)}\sim \left( \frac{1}{p^{2}} \right)^{\frac{1}{2}} \times \mbox{finite}, \label{fusk}
\end{equation}
a $\q$-independent result. By setting $\q=0$ the second term of (\ref{hurra}) will produce a term proportional to $\d^{0a}\d^{0b}$, (\ref{fusk}). This term combines with the first part of (\ref{hurra}) to reproduce the result of the commutative $SU(N)$ model.

\section{Conclusion and Outlook}

In this short note we have demonstrated the finiteness of the $\q$-deformed Chern-Simons theory and its $\q_{\m\n}$-independence at the one-loop level.\\
Additionally, the obtained results are in agreement with the restrictions coming from the VSUSY. In the limit $\q_{\m\n}\rightarrow 0$ one gets the old results of \cite{Guadagnini:1989kr}. Furthermore, since no radiative corrections are present, the model is regular in this limit (see also \cite{Matusis:2000jf} and \cite{Minwalla:1999px}).\\
In a forthcoming paper we will discuss further topological field models of Schwarz-type and Witten-type.

\section{Acknowledgements}
It is our pleasure to thank H. Grosse, H. H\"uffel, R. Wulkenhaar and C. Rupp for helpful discussions in a weekly seminar.

\end{document}